\documentclass[11pt]{article}
\usepackage{amsmath,amssymb}

\makeatletter
\@addtoreset{equation}{section}
\makeatother

\def\ben{\begin{equation}}
\def\een{\end{equation}}

 \let\m=\mu \let\n=\nu  \let\p=\pi

\let\X=\Xi     
\let\C=\Chi

\def\nn{\nonumber} \def\bd{\begin{document}} \def\ed{\end{document}}
\def\ds{\documentstyle} \let\fr=\frac \let\bl=\bigl \let\br=\bigr
\let\Br=\Bigr \let\Bl=\Bigl
\let\bm=\bibitem
\let\na=\nabla
\let\pa=\partial \let\ov=\overline
\newcommand{\be}{\begin{equation}}
\newcommand{\ee}{\end{equation}}
\def\ba{\begin{array}}
\def\ea{\end{array}}
\def\ft#1#2{{\textstyle{\frac{\scriptstyle #1}{\scriptstyle #2} } }}
\def\fft#1#2{{\frac{#1}{#2}}}
\def\del{\partial}
\def\vp{\varphi}
\def\sst#1{{\scriptscriptstyle #1}}
\def\oneone{\rlap 1\mkern4mu{\rm l}}
\def\td{\tilde}
\def\wtd{\widetilde}
\def\ie{{\it i.e.\ }}
\def\dalemb#1#2{{\vbox{\hrule height .#2pt
        \hbox{\vrule width.#2pt height#1pt \kern#1pt
                \vrule width.#2pt}
        \hrule height.#2pt}}}
\def\square{\mathord{\dalemb{6.8}{7}\hbox{\hskip1pt}}}
\newcommand{\ho}[1]{$\, ^{#1}$}
\newcommand{\hoch}[1]{$\, ^{#1}$}
\newcommand{\bea}{\setlength\arraycolsep{2pt} \begin{eqnarray}}
\newcommand{\eea}{\end{eqnarray}}
\newcommand{\ra}{\rightarrow}
\newcommand{\lra}{\longrightarrow}
\newcommand{\Lra}{\Leftrightarrow}
\newcommand{\bp}{\tilde \beta^\prime}
\newcommand{\tr}{{\rm tr} }
\newcommand{\Tr}{{\rm Tr} }
\def\0{{\sst{(0)}}}
\def\1{{\sst{(1)}}}
\def\2{{\sst{(2)}}}
\def\3{{\sst{(3)}}}
\def\4{{\sst{(4)}}}
\def\5{{\sst{(5)}}}
\def\6{{\sst{(6)}}}
\def\7{{\sst{(7)}}}
\def\8{{\sst{(8)}}}
\def\m{{\sst{(m)}}}
\def\n{{\sst{(n)}}}
\def\cA{{{\cal A}}}
\def\cB{{{\cal B}}}
\def\cF{{{\cal F}}}
\def\cG{{{\cal G}}}
\def\cH{{{\cal H}}}
\def\tV{\widetilde V}
\def\tW{\widetilde W}
\def\tH{\widetilde H}
\def\tE{\widetilde E}
\def\tF{\widetilde F}
\def\tA{\widetilde A}
\def\im{{{\rm i}}}
\def\tY{{{\wtd Y}}}
\def\ep{{\epsilon}}
\def\vep{{\varepsilon}}
\def\bD{{{\bar D}}}
\def\R{{{\mathbb R}}}
\def\C{{{\mathbb C}}}
\def\H{{{\mathbb H}}}
\def\CP{{{\mathbb C}{\mathbb P}}}
\def\RP{{{\mathbb R}{\mathbb P}}}
\def\Z{{{\mathbb Z}}}
\def\bA{{{\mathbb A}}}
\def\bB{{{\mathbb B}}}
\def\bC{{{\mathbb C}}}
\def\bD{{{\mathbb D}}}
\def\bE{{{\mathbb E}}}
\def\bZ{{{\mathbb Z}}}
\def\Re{{{\frak{Re}}}}
\def\Im{{{\frak{Im}}}}
\def\cosec{{\,\hbox{cosec}\,}}
\def\Gm{{\Gamma_{\!\! -}}}
\def\Gp{{\Gamma_{\!\! +}}}
\def\stan{{standard }}
\def\nonstan{{supernumerary }}
\def\p{{\partial}}
\def\kdel#1{{\fft{\del}{\del#1}}}

\def\bog{{Bogomolny }}
\def\om{{\omega}}

\newcommand{\nnr}{\nonumber \\}
\newcommand{\pd}{\partial}
\newcommand{\ud}{\textrm{d}}
\newcommand{\dTH}{T^{\prime \, 0}_\textrm{H}}
\newcommand{\dOi}{\Omega^{\prime \, 0}_i}

\newcommand{\tamphys}{\it George and Cynthia Woods Mitchell  Institute
for Fundamental Physics and Astronomy,\\
Texas A\&M University, College Station, TX 77843-4242, USA}

\newcommand{\auth}{
H. L\"u\hoch{\dagger\star}, 
Jianwei Mei\hoch{\dagger}, C.N. Pope\hoch{\dagger,\ddagger}
and Justin F. V\'azquez-Poritz\hoch{\diamondsuit}
}

\begin{document}

\begin{flushright}
\hfill{
MIFP-09-02\ \ \ \ \ \ \ \  }\\
\end{flushright}

\begin{center}
{\large {\bf Extremal Static AdS Black Hole/CFT Correspondence\\
in Gauged Supergravities}}

\vspace{15pt}
\auth

\vspace{10pt}
\hoch{\dagger}{\tamphys}

\vspace{10pt}


\hoch{\star}{\it Division of Applied Mathematics and Theoretical
Physics,\\
China Institute for Advanced Study,\\
Central University of Finance and Economics, Beijing, 100081, China
}

\vspace{10pt}

\hoch{\ddagger}{\it  DAMTP, Centre for Mathematical Sciences,
 Cambridge University,\\  Wilberforce Road, Cambridge CB3 OWA, UK}

\vspace{10pt}

\hoch{\diamondsuit}{\it Physics Department,\\ 
New York City College of Technology,\\ the City University of New York, 
Brooklyn, NY 11201, USA}

\vspace{30pt}

\underline{ABSTRACT}
\end{center}

   A recently proposed holographic duality allows the Bekenstein-Hawking
entropy of extremal rotating black holes to be calculated microscopically,
by applying the Cardy formula to the two-dimensional chiral CFTs associated
with certain reparameterisations of azimuthal angular coordinates in
the solutions.  The central charges are proportional to the angular
momenta of the black hole, and so the method degenerates in the case of
static (non-rotating) black holes.  We show that the method can be extended
to encompass such charged static extremal AdS black holes by using consistent
Kaluza-Klein sphere reduction ans\"atze to lift them to exact solutions in
the low-energy limits of 
string theory or M-theory, where the electric charges become reinterpreted
as angular momenta associated with internal rotations in the reduction sphere.
We illustrate the procedure for the examples of extremal charged static
AdS black holes in four, five, six and seven dimensions.

\vspace{15pt}

\thispagestyle{empty}




\newpage
\section{Introduction}

The first computation of black hole entropy by counting microstates was
performed for the case of five-dimensional extremal black holes 
\cite{stromvafa}.
This was subsequently extended to a variety of other examples, including
black holes in four dimensions \cite{maldstrom}.
In this approach, the microscopic black hole states are equated with
BPS D-brane states, and so it seemingly depends on features  
both of supersymmetry and of string theory.

On the other hand, it had already been known for quite some time that
any consistent quantum theory of gravity on three-dimensional anti-de
Sitter spacetime AdS$_3$ is equivalent to a two-dimensional conformal
field theory, owing to the fact that the asymptotic symmetry group of
AdS$_3$ is generated by (two copies of) the Virasoro algebra \cite{brown}.
Combining the evaluation of the central charge 
with Cardy's formula for the asymptotic
growth of states gives rise to a microscopic computation of entropy for
black holes whose near-horizon geometry is locally AdS$_3$, such as the
BTZ black hole. When the aforementioned extremal black holes in the dilute
gas approximation \cite{dilutegas} are
embedded in one extra dimension, the near-horizon geometry becomes
a direct product of a BTZ black hole and a sphere.   The dilute
gas approximation can be applied also to a rotating black hole
\cite{cveticlarsen}.  This enables one to compute the statistical
entropy without invoking supersymmetry or string theory \cite{strom}.

Another holographic correspondence, namely AdS/CFT \cite{agmoo},
has been extensively applied to a variety of black holes in
gauged supergravity.  The most concrete tests of AdS/CFT
have been performed in cases for which the supergravity backgrounds are
supersymmetric, although this duality may extend to non-supersymmetric
backgrounds also.  However, a {\it static} AdS black hole in the supersymmetric
limit suffers from a naked singularity.  This singularity can be avoided by
adding rotation.  Supersymmetric rotating AdS black holes
in five dimensions were obtained in \cite{real1,cclp,cclp2,real2},
and subsequently, a boundary free-fermion approximation was
used in order to obtain a microscopic evaluation of the entropy, 
up to a numerical factor of order unity
\cite{kmmr}.

Recently, a proposal was made for a new holographic duality
between four-dimensional extremal Kerr black holes and a two-dimensional
chiral conformal field theory \cite{guhasost}, which
extends the previous approaches of \cite{carlip,sol1,par1,par2,hot}.
This proposed duality
is motivated by the observation that the Cardy formula yields an entropy 
for the
CFT which is in precise agreement with the Bekenstein-Hawking entropy
of the extremal black hole. This proposed Kerr/CFT correspondence was 
subsequently extended
to asymptotically flat and asymptotically AdS 
black holes with multiple angular momenta in diverse dimensions
\cite{lumeipope}, where it was found that there is a chiral
two-dimensional CFT associated with each independent rotation. Additional
generalizations include Kaluza-Klein black holes \cite{Azeyanagi:2008kb},
the Kerr-Newman-AdS black hole \cite{hamunist}, NS5-branes
\cite{Nakayama:2008kg}, large classes of rotating black holes in gauged and
ungauged supergravities \cite{cclpbhcft}, and several other types
of rotating black holes \cite{isono,azeyanagi,peng,chen}.

The calculation of the central charge is similar to the approach taken
in \cite{strom}, which is based on \cite{brown,Barnich:2001jy}. 
Namely, one considers the
asymptotic symmetry generators associated with a class of perturbations
around the near-horizon Kerr geometry that obey suitably-chosen
boundary conditions. This new method of computing microscopic entropy
does not require the black hole to be supersymmetric or to be embedded
within string theory, although it must be extremal. The rotation of the
black hole plays a vital role, however, 
since the central charge is proportional
to the angular momentum. In fact, the thermodynamic description breaks
down for static black hole, in the sense that the central charge vanishes
and the temperature diverges. 
 
   Since one can nonetheless use the Cardy formula to obtain the
correct entropy for static black holes by taking a limit of rotating 
black holes, this suggests that there may be an alternative strictly static
description that is not singular.  In this paper we present such a description,
based upon the observation that static charged black holes in many gauged 
supergravities can be lifted, by means of consistent Kaluza-Klein
reduction formulae derived in \cite{10author}, 
to become solutions in the ten or eleven-dimensional
supergravities that arise as the low-energy limits of string theory 
or M-theory.
The electric charges of the static black holes acquire the interpretation
of rotations in the internal (spherical) dimensions, after the lifting
has been performed.\footnote{A similar idea was discussed in \cite{hamunist},
where it was argued that the extremal four-dimensional Kerr-Newman-AdS 
black hole could be viewed as a neutral five-dimensional configuration with a
rotation in the fifth dimension.  However, no consistent Kaluza-Klein
reduction from five dimensions can give rise to the four-dimensional
Einstein-Maxwell theory with cosmological constant, and so
the four-dimensional Kerr-Newman-AdS black hole cannot be lifted to an
actual neutral solution of any five-dimensional theory.  
Thus the example considered in
\cite{hamunist} is perhaps somewhat heuristic in nature.}
The procedure developed in \cite{guhasost} can then
be applied to the lifted solutions, with the Cardy formula for the 
entropies of the dual CFTs associated with the internal rotations
giving a microscopic derivation of the Bekenstein-Hawking entropy of
the original lower-dimensional extremal static AdS black hole. 
We apply this procedure to a variety of extremal static $U(1)$
charged AdS black holes in gauged supergravities in diverse dimensions.
In particular, we consider five-dimensional 3-charge AdS black holes
\cite{bcs}, four-dimensional 4-charge AdS black holes \cite{duffliu},
seven-dimensional 2-charge AdS black holes \cite{10author} and
six-dimensional single-charge AdS black holes \cite{romanstoromans}.
We show
that the microscopic entropy matches perfectly with the Bekenstein-Hawking
entropy for each of these cases.

\section{Central charges of near-extremal metrics}

        In this section, we consider a large class of metrics that
can be viewed as  ${\cal M}$ bundles over AdS$_2$, where ${\cal M}$
can be any smooth manifold.  The general form of the metrics is
given by
\bea
ds^2 &=& A \left( -(1 + r^2) d\tau^2 + \fft{dr^2}{1 + r^2} \right) +
h_{\alpha\beta}\, dy^\alpha\, dy^\beta + \tilde
g_{ij}\, \tilde e^i \tilde e^j\,,\nn\\
\tilde e^i &=& d\phi_i + k_i r\, d\tau \,,\label{dgenh}
\eea
where
$A$, $h_{\alpha\beta}$ and $\td g_{ij}$ can be functions of the
coordinates $y^\alpha$, and $k_i$'s are constants.  Let us assume
that the index $i$ runs from 1 to $n$.  There exist $n$ commuting
diffeomorphisms that preserve the boundary structure at
$r\rightarrow \infty$, namely
\be
\zeta^i_m = - e^{-\im m \phi_i}\, \fft{\del}{\del\phi_i} - \im m\, r\,
   e^{-\im m \phi_i}\, \fft{\del}{\del r}\,,\qquad
i=1,\ldots n\,.
\ee
These diffeomorphisms generate $n$ commuting Virasoro algebras.
The central charges $c_i$ in these Virasoro algebras, at the level of
Dirac brackets
of the associated charges $Q^i_\n= 1/(8\pi)\int_{\del\Sigma} k^i_\n$,
can be calculated in the manner described in \cite{Barnich:2001jy},
namely from the $m^3$ terms in the expressions
\be
\fft1{8\pi}\, \int_{\del\Sigma}
k_{\zeta^i_\m}[{\cal L}_{\zeta^i_\sst{(-m)}}g,g]=
 -\fft{\im}{12} (m^3 + \alpha m) c_i\,,\label{central}
\ee
where
\bea
k_\zeta[h,g] &=&\ft12 \Big[ \zeta_\nu\nabla_\mu h
  - \zeta_\nu \nabla_\sigma h_\mu{}^\sigma +
  \zeta_\sigma\nabla_\nu h_\mu{}^\sigma + \ft12 h \nabla_\nu\zeta_\mu
- h_\nu{}^\sigma \nabla_\sigma\zeta_\mu\nn\\
&&\qquad\qquad\qquad + \ft12 h_{\nu\sigma}
(\nabla_\mu\zeta^\sigma + \nabla_\sigma\zeta_\mu)\Big] \, {*(dx^\mu\wedge
dx^\nu)}\,.\label{kdef}
\eea
Taking $g_{\mu\nu}$ to be given by (\ref{dgenh}), we find that the
central charges are \footnote{The ansatz (\ref{dgenh}) is slightly
more general than the one originally presented in \cite{cclpbhcft}, in that
the metric contribution
$h_{\alpha\beta}\, dy^\alpha\, dy^\beta$ associated with the
coordinates $y^\alpha$ is not restricted to being diagonal.}
\bea
c_i &=&  \fft{3k_i {\cal A}}{2\pi}\,,\qquad
{\cal A}= \int\sqrt{h\, \td g}\, d^p y\,
         \int \prod_i d\phi_i\,,\label{cent1}
\eea
where ${\cal A}$ is the volume of the manifold ${\cal M}$.

       The structures of the near-horizon geometries of extremal black holes
were extensively studied previously \cite{kun1,kun2}, and were found
to be encompassed within the general
form of (\ref{dgenh}).  It follows that the integrals in
(\ref{cent1}) can be identified with the entropies of the extremal
black holes.  In \cite{cclpbhcft},
it was shown by examining a wide class of rotating black holes that
the constants $k_i$ are given by
\be
k_i = \fft{1}{2\pi T_i}\,,\label{frolov}
\ee
where $T_i$ is the associated Frolov-Thorne temperature on the horizon
\cite{frotho}.
It follows, therefore, that the central charge is related to the 
Bekenstein-Hawking entropy $S_{BH}$ by
\be 
c_i = \fft{6 k_i\, S_{BH}}{\pi}\,.\label{cent2}
\ee
This shows that any near-horizon geometry of the form (\ref{dgenh})
will have the property that the microscopic entropy for the $i$'th
CFT associated with reparameterisations of $\phi_i$, calculated
using the Cardy formula
\be
S_{BH} = \fft{\pi^2}{3}\, c_i\, T_i\,, \qquad \hbox{for each $i$}
\label{cardy}
\ee
will agree precisely with the Bekenstein-Hawking entropy $S_{BH}$.  

   It is perhaps worth emphasising at this point that the agreement 
between the microscopic calculation of the dual CFT entropy and the
Bekenstein-Hawking calculation of the black hole entropy would 
break down if the constants $k_i$ in (\ref{dgenh}) were equal to
zero, which would be the case for static black holes.  As can be seen
from (\ref{frolov}), the Frolov-Thorne temperature $T_i$ would be infinite,
while, from (\ref{cent2}), the central charge $c_i$ would vanish.  One
could still obtain the proper finite and non-zero result for the
entropy of a static extremal black hole, using (\ref{cardy}), by taking a 
static limit of rotating black
holes.  But if instead one starts form a black hole that is exactly static,
then (\ref{cardy}) cannot be used.  This problem is circumvented by lifting
the static black holes to higher dimensions, as we shall describe in the
remainder of this paper.

\section{Five-dimensional 3-charge AdS black holes}

The maximal gauged supergravity in $D=5$ has $SO(6)$ gauge
symmetry.  The Cartan subgroup is $U(1)^3$.   The five-dimensional
three-charge static AdS black hole solution was constructed in
\cite{bcs}.  We adopt the convention of \cite{10author}, and the
solution is given by
\bea\label{bh5}
ds_5^2 &=& -{\cal H}^{-2/3} f\ d\hat t^2+{\cal H}^{1/3}
(f^{-1} d\hat r^2+\hat r^2 d\Omega_{3,\epsilon}^2)\,,\nn\\
X_i &=& H_i^{-1} {\cal H}^{1/3}\,,
\qquad A_{(1)}^i=\Phi_i\, d\hat t\,,\qquad
\Phi_i = -(1-H_i^{-1}) \alpha_i \,,\nn\\
f &=& \epsilon-\fft{\mu}{\hat r^2}+g^2\hat r^2 {\cal H}\,,
\qquad {\cal H}=H_1H_2H_3\,,\qquad
H_i=1+\fft{\ell_i^2}{\hat r^2}\,,\cr
\alpha_i &=&\fft{\sqrt{1 + \epsilon \sinh^2\beta_i}}{
\sinh\beta_i}\,, \qquad \ell_i^2 = \mu\sinh^2\beta_i\,,
\label{d5bh}
\eea
where $d\Omega_{3,\epsilon}^2$ is the unit metric for $S^3$, $T^3$ or
$H^3$ for $\epsilon=1, 0$ or $-1$, respectively.  If all the charge
parameters $\beta_i$ are set equal, the solution becomes the
five-dimensional Reissner-Nordstr\"om AdS black hole.
The outer horizon is located at $\hat r=r_+$, which is the largest root of $f$.
The temperature and entropy are given by
\be
T_H = \fft{f'(r_+)}{4\pi \sqrt{{\cal H}(r_+)}}\,,\qquad
S = \ft14 r_+^3\, \omega_{3,\epsilon}\,  \sqrt{{\cal H}(r_+)}\,,
\ee
where $\omega_{3,\epsilon}$ is the volume for the
$d\Omega_{3,\epsilon}^2$.  The extremal limit is obtained
when the function $f$ has a double zero, $r=r_0$.  This can
be achieved by choosing parameters such that\footnote{It should be emphasised
that the {\it extremal limit} is quite different from the {\it BPS limit},
which is obtained by sending $\mu$ to zero.  Whilst the extremal limit 
is non-singular, the BPS limit has a naked singularity at $r=0$.  Similar
remarks apply to the static AdS black holes in other dimensions that
we discuss in subsequent sections.}
\be
\mu = \fft{g^2}{r_0^2} \left(2\ell_{123}^2 +
r_0^2 \sum_{i<j}\ell_{ij}^2- r_0^6\right)\,,\qquad
\epsilon = \fft{g^2}{r_0^4} \left(\ell_{123}^2 -
r_0^4 \sum_{i<j}\ell_{ij}^2- 2 r_0^6\right)\,.
\ee
In this paper, we define $\ell_{i_1\ldots i_n}=\ell_{i_1} \cdots
\ell_{i_n}$. 
In this extremal limit, the temperature vanishes, but the entropy
is non-vanishing, given by
\be
S_0= \fft1{4} r_0^3\, \omega_{3,\epsilon}\,  \sqrt{{\cal H}_0}\,,
\label{d5exentropy}
\ee
where ${\cal H}_0\equiv {\cal H}(r_0)$. 
In the extremal limit, the near-horizon geometry of the
black hole is the direct product AdS$_2\times S^3$.  
There exists a decoupling
limit in which the near-horizon geometry becomes a solution in
its own right.  To see this, we note that in the near horizon,
the function $f$ can be expanded as
\be
f=(\hat r-r_0)^2 V\,,\qquad
V=\ft12 f''(r_0)= \fft{4g^2}{r_0^6} (\ell_{123}^2 + r_0^6)\,.
\ee
Making the coordinate transformation
\be\label{trans}
\hat r=r_0 (1 + \lambda \rho)\,,\qquad
\hat t=\fft{\sqrt{{\cal H}_0}}{\lambda r_0 V} t\,,
\ee
and then sending the constant parameter $\lambda\rightarrow 0$,
the solution becomes
\bea
ds_5^2 &=& \fft{{\cal H}_0^{1/3}}{V} \left(-\rho^2 dt^2 +
 \fft{d\rho^2}{\rho^2} \right) + r_0^2 {\cal H}_0^{1/3}
d\Omega_{3,\epsilon}^2\,,\cr
X_i^0 &=& \fft{{\cal H}_0^{1/3}}{H_i(r_0)}\,,\qquad
A_\1^i = \fft{k_i \rho}{g} dt\,,\label{ads2s3}
\eea
where the constant $k_i$ is given by
\be
k_i = \fft{1}{2\pi T_i}\,,\qquad T_i =- \fft{T_H'(r_0)}{g\Phi'(r_0)}=
\fft{g (r_0^2 + \ell_i^2)^2 (\ell_{123}^2 + r_0^6)}
{\pi r_0^7 \alpha_i\ell_i^2\sqrt{{\cal H}_0}}\,.\label{kiti}
\ee
Note that we have extracted the pure constant divergent terms of
$A_\1^i$ in (\ref{ads2s3}) as pure gauge.  

   The metric (\ref{ads2s3}) can be recast in terms of global AdS$_2$
coordinates $(\tau,r)$ rather than the Poincar\'e patch coordinates 
$(t,\rho)$ by means of the transformations
\be
\rho= r + \sqrt{1+r^2}\, \cos\tau\,,\qquad
t= \fft{\sqrt{1+r^2}\, \sin\tau}{r+ \sqrt{1+r^2}\, \cos\tau}\,.
\ee
After absorbing an exact form into the potential $A_\1^i$ by means of a
gauge transformation, the solution (\ref{ads2s3}) becomes
\bea
ds_5^2 &=& \fft{{\cal H}_0^{1/3}}{V} \left(-(1+r^2) d\tau^2 +
 \fft{dr^2}{1+r^2} \right) + r_0^2 {\cal H}_0^{1/3}
d\Omega_{3,\epsilon}^2\,,\cr
X_i^0 &=& \fft{{\cal H}_0^{1/3}}{H_i(r_0)}\,,\qquad
A_\1^i = \fft{k_i r}{g} d\tau \,,\label{ads2s32}
\eea

     We can now lift the solution back to $D=10$, using the
reduction ansatz given in \cite{10author}, finding that
the metric is given by
\bea
ds_{10}^2 &=& \sqrt{\Delta} ds_5^2 +
\fft{1}{g^2 \sqrt{\Delta}} \sum_{i=1}^3 X_i^{-1} \left(
d\mu_i^2 + \mu_i^2 (d\hat \phi + g\, A_\1^i)^2\right)\,,\cr
\Delta &=& {\cal H}^{1/3} \sum_{i=1}^3 \fft{\mu_i^2}{H_i}\,,
\qquad \sum_{i=1}^3 \mu_i^2=1\,.\label{d5red}
\eea
The near-horizon geometry of the black hole in the extremal limit
is given by
\bea
ds_{10}^2 &=& \fft{\sqrt{\Delta_0}\, {\cal H}_0^{1/3}}{V}
\left[-(1+r^2) d\tau^2 + \fft{dr^2}{1+r^2} +
V r_0^2\, d\Omega_{3,\epsilon}^2\right]\cr
&&+\fft{1}{g^2\sqrt{\Delta_0}} \sum_{i=1}^3 (X_i^0)^{-1}
\left(d\mu_i^2 + \mu_i^2 (d\phi_i + k_i r\, d\tau)^2\right)\,,
\label{nh1}
\eea
where $\Delta_0 = \Delta(r_0)$.  The metric can be viewed
as a warped $S^3\times S^5$ bundle over AdS$_2$, with the fibre
lying only in the $S^5$ directions.  The volume of the
warped $S^3\times S^5$ is given by
\be
{\cal A} = g^{-5}\, r_0^3 \sqrt{{\cal H}_0}\, \omega_{3,e} \omega_5
\,.
\ee
where $\omega_5$ is the volume for the unit $S^5$.
The near-horizon metric (\ref{nh1}) is clearly contained within the
general ansatz (\ref{dgenh}); we may take
\bea
h_{\alpha\beta}\, dy^\alpha\, d y^\beta &=&
 \sqrt{\Delta_0}\, {\cal H}_0^{1/3}\, r_0^2\, d\Omega_{3,\ep}^2 +
   \fft1{g^2\, \sqrt{\Delta_0}}\, \sum_{i=1}^3\, (X_i^0)^{-1}\, d\mu_i^2\,,
\nn\\
\td g_{ij} &=& \fft{\mu_i^2}{g^2\, \sqrt{\Delta_0}\, X_i^0}\, \delta_{ij}\,.
\eea
It therefore follows from the general discussion given earlier that
the central charge of the $i$'th Virasoro symmetry associated with
reparameterisations of $\phi_i$ is given by
\be
c_i = \fft{3k_i {\cal A}}{2\pi G_{10}} =
\fft{6k_i S_0}{\pi}\,,
\ee
where $S_0$ is the $D=5$ black hole entropy given in (\ref{d5exentropy}).
Here we have temporarily restored Newton's constant, which enters in the
denominator of the Hawking entropy, $S={\cal A}/(4G)$, and which we
normally set to unity, in order to discuss the relation between the
entropy in five dimensions and in ten dimensions.
This follows by noting that the 
Kaluza-Klein reduction ansatz given in \cite{10author}
implies that the Newton constants in ten and five dimensions are related
by $G_{10}=g^{-5} \omega_5\, G_5$.  Since the horizon areas are also
related by ${\cal A}_{10} = g^{-5} \omega_5\, {\cal A}_5$, it follows that
the ten-dimensional and five-dimensional entropies are equal.  An analogous
result holds in all the examples in other dimensions that we discuss in
subsequent sections.

   From the reduction ansatz (\ref{d5red}), we see that the electric potential
$\Phi_i$ is related to the angular velocities $\Omega_i$ of
the azimuthal angles $\phi_i$ in $D=10$ by
\be
\Omega_i = g\, \Phi_i\,.
\ee
It follows that $T_i$, given in (\ref{kiti}), can be identified
as the Frolov-Thorne temperature, and therefore that the entropy
calculated using the Cardy formula (\ref{cardy}),
will agree precisely with the Bekenstein-Hawking entropy of the
the extremal five-dimensional static AdS black
hole (\ref{d5bh}).

       It is of interest to note that we can perform a Kaluza-Klein
reduction on the 3-manifold $d\Omega_{3\epsilon}^2$.  The resulting
solution becomes a rotating black hole in $D=7$, and the extremal 
black hole/CFT correspondence continues to hold.

\section{Four-dimensional 4-charge AdS black holes}

The maximum gauged supergravity in $D=4$ has $SO(8)$ gauge
group, whose Cartan subgroup is $U(1)^4$.   The four-charge
static AdS black hole was constructed in \cite{duffliu,sabra}

Following the convention of \cite{10author}, the four-dimensional
4-charge AdS black hole solution is given by
\bea\label{bh4}
ds_4^2 &=& -{\cal H}^{-1/2} f\ d\hat t^2+{\cal H}^{1/2}
(f^{-1} d\hat r^2+\hat r^2 d\Omega_{2,\epsilon}^2)\,,\nn\\
X_i &=& H_i^{-1} {\cal H}^{1/4}\,,
\qquad A_{(1)}^i=\Phi_i\, d\hat t\,,\qquad
\Phi_i = -(1-H_i^{-1}) \alpha_i \,,\nn\\
f &=& \epsilon-\fft{\mu}{\hat r}+4g^2\hat r^2 {\cal H}\,,
\qquad {\cal H}=H_1H_2H_3H_4\,,\qquad
H_i=1+\fft{\ell_i}{\hat r}\,,\cr
\alpha_i &=&\fft{\sqrt{1 + \epsilon \sinh^2\beta_i}}{
\sinh\beta_i}\,, \qquad \ell_i = \mu\sinh^2\beta_i\,,
\eea
where $d\Omega_{2,\epsilon}^2$ is the unit metric for $S^2$, $T^2$ or
$H^2$ for $\epsilon=1, 0$ or $-1$, respectively.  If the charge
parameters $\beta_i$ are set equal, the solution becomes the standard
Reissner-Nordstr\"om AdS black hole.
The outer horizon is located at $\hat r=r_+$, 
which is the largest root of $f$. The temperature and entropy are given by
\be
T_H = \fft{f'(r_+)}{4\pi \sqrt{{\cal H}(r_+)}}\,,\qquad
S = \ft14 r_+^2\, \omega_{2,\epsilon}\,  \sqrt{{\cal H}(r_+)}\,,
\ee
where $\omega_{2,\epsilon}$ is the volume for the
$d\Omega_{2,\epsilon}^2$.  The extremal limit is obtained
when the function $f$ has a double zero, $r=r_0$.  This can
be achieved by choosing parameters such that
\bea
\mu &=& \fft{4g^2}{r_0} \left(2\ell_{1234}+r_0\sum_{i<j<k}\ell_{ijk}
-r_0^3\sum_{i}\ell_{i}-2r_0^4\right)\,,\cr
\epsilon &=& \fft{4g^2}{r_0^2} \left(\ell_{1234} -
r_0^2 \sum_{i<j}\ell_{ij}-2r_0^3
\sum_{i}\ell_{i} -3 r_0^4\right)\,.
\eea
In this extremal limit, the temperature vanishes, but the entropy
is non-vanishing, given by
\be
S_0= \fft1{4} r_0^2 \omega_{2,\epsilon}\,  \sqrt{{\cal H}_0}\,,
\label{d4exentropy}
\ee
where ${\cal H}_0\equiv {\cal H}(r_0)$.
In the extremal limit, the near-horizon geometry of the
black hole is AdS$_2\times S^2$.  There exists a decoupling
limit that the near-horizon geometry is a solution on
its own.  To see this, we note that in the near horizon,
the function $f$ can be expanded as
\be
f=(\hat r-r_0)^2 V\,,\qquad
V=\ft12 f''(r_0)= \fft{4g^2}{r_0^4} \left(\ell_{1234}+ 
r_0^3\sum_{i}\ell_{i} + 3r_0^4 \right)\,.
\ee
Making the coordinate transformation (\ref{trans})
and then sending the constant parameter $\lambda\rightarrow 0$,
the solution in global coordinates becomes
\bea
ds_4^2 &=& \fft{{\cal H}_0^{1/2}}{V} \left(-(1 + r^2) d\tau^2 +
\fft{dr^2}{1 + r^2} \right) + r_0^2 {\cal H}_0^{1/2}
d\Omega_{2,\epsilon}^2\,,\cr
X_i &=& \fft{{\cal H}_0^{1/4}}{H_i(r_0)}\,,\qquad
A_\1^i = \fft{k_i r}{g} d\tau\,,\label{ads2s2}
\eea
where the constant $k_i$ is given by
\be
k_i = \fft{1}{2\pi T_i}\,,\qquad
T_i = - \fft{T_H'(r_0)}{g\Phi_i'(r_0)}\,.
\ee

     Using the reduction ansatz given in \cite{10author}, we can
now lift the solution back to $D=11$.  The metric is given by
\bea
ds_{11}^2 &=& \fft{\Delta_0^{2/3} {\cal H}_0^{1/2}}{V} \left(
-(1+r^2) d\tau^2 + \fft{dr^2}{1+r^2} + V\, r_0^2
d\Omega_{2,\epsilon}^2\right)\cr
&& +
\fft{1}{g^2\Delta_0^{1/3}}\sum_{i=1}^4 \fft{1}{X_i^0}
\left(d\mu_i^2 + \mu_i^2 (d\phi_i + k_i\,r d\tau)^2\right)\,,\cr
\Delta_0 &=& \sum_{i=1}^4 X_i^0 \mu_i^2\,,\qquad
X_i^0 = X_i(r_0)\,.
\eea
Thus we see that the metric fits the general ansatz (\ref{dgenh}),
and $T_i$ can be identified as the Frolov-Thorne temperature.
Following the same discussion in the previous section, the microscopic
entropy for the $i$'th CFT associated with the reparameterisations
of $\phi_i$, calculated using the Cardy formula will agree
precisely with the Bekenstein-Hawking entropy.

\section{Seven-dimensional 2-charge AdS black holes}

The maximal gauged supergravity in $D=7$ has $SO(5)$ gauge symmetry,
whose Cartan subgroup is $U(1)^2$.
The seven-dimensional 2-charge AdS black hole solution is given by
\cite{10author}
\bea\label{bh7}
ds_7^2 &=& -{\cal H}^{-4/5} f\ d\hat t^2+{\cal H}^{1/5}
(f^{-1} d\hat r^2+\hat r^2 d\Omega_{5,\epsilon}^2)\,,\nn\\
X_i &=& H_i^{-1} {\cal H}^{2/5}\,,
\qquad A_{(1)}^i=\Phi_i\, d\hat t\,,\qquad
\Phi_i = -(1-H_i^{-1}) \alpha_i \,,\nn\\
f &=& \epsilon-\fft{\mu}{\hat r^4}+\ft14 g^2\hat r^2 {\cal H}\,,
\qquad {\cal H}=H_1H_2\,,\qquad
H_i=1+\fft{\ell_i^4}{\hat r^4}\,,\cr
\alpha_i &=&\fft{\sqrt{1 + \epsilon \sinh^2\beta_i}}{
\sinh\beta_i}\,, \qquad \ell_i^4 = \mu\sinh^2\beta_i\,,
\eea
where $d\Omega_{5,\epsilon}^2$ is the unit metric for $S^5$, $T^5$ or
$H^5$ for $\epsilon=1, 0$ or $-1$, respectively.
The horizon is at $\hat r=r_+$, which is the largest root of $f$.
The temperature and entropy are given by
\be
T_H = \fft{f'(r_+)}{4\pi \sqrt{{\cal H}(r_+)}}\,,\qquad
S = \ft14 r_+^5\, \omega_{5,\epsilon}\,  \sqrt{{\cal H}(r_+)}\,,
\ee
where $\omega_{5,\epsilon}$ is the volume for the
$d\Omega_{5,\epsilon}^2$.  The extremal limit is obtained
when the function $f$ has a double zero, $r=r_0$.  This can
be achieved by choosing parameters such that
\be
\mu = \fft{g^2}{8r_0^2} \left(3\ell_{12}^4 +
r_0^4 (\ell_1^4 + \ell_2^4) - r_0^8\right)\,,\qquad
\epsilon = -\fft{g^2}{8r_0^6} \left(\ell_{12}^4 - r_0^4
(\ell_1^4+\ell_2^4) - 3 r_0^8\right)\,.
\ee
In this extremal limit, the temperature vanishes, but the entropy
is non-vanishing, given by
\be
S_0= \ft14 r_0^5 \omega_{5,\epsilon}\,  \sqrt{{\cal H}_0}\,,
\label{d7exentropy}
\ee
where ${\cal H}_0\equiv {\cal H}(r_0)$.
In the extremal limit, the near-horizon geometry of the
black hole is AdS$_2\times S^5$.  There exists a decoupling
limit that the near-horizon geometry is a solution on
its own.  To see this, we note that in the near horizon,
the function $f$ can be expanded as
\be
f=(\hat r-r_0)^2 V\,,\qquad
V=\ft12 f''(r_0)= \fft{g^2}{2r_0^8} (3\ell_{12}^4 -
r_0^4 (\ell_1^4 + \ell_2^4) +3 r_0^8)\,.
\ee
Making the coordinate transformation (\ref{trans})
and then sending the constant parameter $\lambda\rightarrow 0$,
the solution in AdS$_2$ global coordinates becomes
\bea
ds_5^2 &=& \fft{{\cal H}_0^{2/5}}{V} \left(- (1+r^2) d\tau^2 +
\fft{dr^2}{1 + r^2}\right) + r_0^2 {\cal H}_0^{2/5}
d\Omega_{5,\epsilon}^2\,,\cr
X_i &=& \fft{{\cal H}_0^{2/5}}{H_i(r_0)}\,,\qquad
A_\1^i = \fft{k_i r}{g} d\tau\,,\label{ads2s5}
\eea
where the constant $k_i$ is given by
\be
k_i = \fft{1}{2\pi T_i}\,,\qquad
T_i = - \fft{T_H'(r_0)}{g\Phi_i'(r_0)}\,.
\ee

      Again we can use the reduction ansatz in \cite{10author} to
lift the solution back to $D=11$.  The metric is given by
\bea
ds_{11}^2 &=& \fft{\Delta_0^{1/3} {\cal H}_0^{2/5}}{V}
\left(-(1+r^2) d\tau^2 + \fft{dr^2}{1+r^2} +
V\, r_0^2 d\Omega_{5,\epsilon}^2\right)\cr
&&+\fft{1}{g^2\Delta_0^{2/3}}\left(
\fft{1}{X_0^0} d\mu_0^2 + \sum_{i=1}^2 \fft{1}{X_i^0} (d\mu_i^2 +
k_i r\, d\tau)^2\right)\,,\cr
\Delta_0&=& \sum_{\alpha=0}^2 X_\alpha^0 \mu_\alpha^2\,,\qquad
X_i^0=X_i(r_0)\,,\qquad X_0^0 = (X_1^0 X_2^0)^{-2}\,.
\eea
Thus we see that the metric fits the general ansatz (\ref{dgenh}),
and $T_i$ can be identified as the Frolov-Thorne temperature.
Following the same discussion as in the previous sections, the microscopic
entropy for the $i$'th CFT associated with the reparameterisations
of $\phi_i$, calculated using the Cardy formula will agree
precisely with the Bekenstein-Hawking entropy.

\section{Six-dimensional single-charge AdS black holes}

       The gauged supergravity in $D=6$ constructed in \cite{romans}
has a $SU(2)$ gauge symmetry.  The $U(1)$ charged AdS black hole
was constructed in \cite{romanstoromans}.   The solution is given by
\bea
ds_6^2 &=& - H^{-3/2} f\, d\hat t^2 + H^{1/2} (f^{-1} d\hat r^2 +
\hat r^2 d\Omega_{4,\epsilon}^2)\,,\cr
X&=&H^{-1/4}\,,\qquad
A_\1 = \Phi\, d\hat t\,,\qquad
\Phi= -\sqrt2 (1-H^{-1}) \alpha\, d\hat t\,,\cr
f&=&\epsilon - \fft{\mu}{\hat r^3} + \ft29 g^2 \hat r^2 H^2\,,\qquad
H=1 + \fft{\ell^3}{\hat r^3}\,,\cr
\alpha &=& \fft{\sqrt{1 + \epsilon\,\sinh^2\beta}}{\sinh\beta}\,,\qquad
\ell^3 = \mu\,\sinh^2\beta\,.
\eea
The temperature and the entropy are given by
\be
T_{H} = \fft{f'(r_+)}{4\pi H(r_+)}\,,\qquad
S=\ft14 r_+^4\, H(r_+) \,\omega_{4,\epsilon}\,,
\ee
where $r_+$ is the largest root of $f$.  The solution becomes
extremal when the parameters satisfy
\be
\mu=\fft{4g^2(2\ell^6 + \ell^3 r_0^3 - r_0^6)}{27r_0}\,,\qquad
\epsilon=\fft{2g^2(\ell^6-4\ell^3 r_0^3 - 5r_0^6)}{27r_0^4}\,.
\ee
In this case, the function $f$ has a double zero at $\hat r=r_0$, and
its expansion near $\hat r=r_0$ is given by
\be
f=(\hat r-r_0)^2 V\,,\qquad
V=\ft12 f''(r_0) = \fft{2g^2 (2\ell^6 - 2\ell^3 r_0^3 + 5r_0^6)}{9r_0^6}\,.
\ee
Using the same method as in the previous examples, we obtain
the near-horizon solution, written in AdS$_2$ global coordinates,
given by
\bea
ds_6^2 &=& \fft{H_0^{1/2}}{V} \left(-(1+r^2) d\tau^2 + \fft{dr^2}{1+r^2}
\right)
 + r_0^2 H_0^{1/2} r_0^2 d\Omega_{4,\epsilon}^2\,,\cr
X_i&=&X_i^0 = H_0^{-1/4}\,,\qquad
A_\1 = \fft{k r}{g} d\tau\,,
\eea
where 
\be
k = \fft{1}{2\pi T}\,,\qquad
T = - \fft{T_H'(r_0)}{g\Phi'(r_0)}\,.
\ee
It was shown in \cite{romanstoromans} that the six-dimensional
gauged supergravity can be obtained by Kaluza-Klein reduction
from the massive type IIA theory \cite{massive} in $D=10$.  The reduction
ansatz has a singular warp factor, associated with the D4/D8 system.
However, the metric in the D4-brane frame is regular \cite{Chong:2004kf}.
Following the reduction ansatz in \cite{romanstoromans}, we have
\bea
ds_{10}^2 &=& \fft{\Delta_0^{2/5} H_0^{1/2}}{V} \left(
-(1+r^2) d\tau^2 + \fft{dr^2}{1+r^2} +
V r_0^2 d\Omega_{4,\epsilon}^2\right)\cr
&& + \fft{\Delta_0^{2/5} X_0^2}{2g^2} d\xi^2 +
\fft{\Delta_0^{3/5}}{2g^2\X_0} \cos^2\xi \left(d\theta^2 +
\sin^2\theta d\phi^2 + (\sigma_3 + 2\, k\, r d\tau)^2\right)\,,\cr
\Delta_0 &=& X_0 \cos^2\xi + X_0^{-3} \sin^2\xi\,,\qquad
X_0=X(r_0)\,,\nn\\
\sigma_3&=& d\psi + \cos\theta\, d\phi\,.
\eea
Thus we see that the metric fits the general ansatz (\ref{dgenh}),
and $T$ can be identified as the Frolov-Thorne temperature.
Following the same discussion in the previous sections, the microscopic
entropy for the CFT associated with the reparameterisations
of $\psi$, calculated using the Cardy formula, will agree
precisely with the Bekenstein-Hawking entropy.

\section{Conclusions}

  In this paper, we have extended the extremal black hole/CFT correspondence 
to static $U(1)$ charged black holes in gauged supergravities in a variety 
of dimensions. The first step was to use a limiting procedure to obtain a 
near-horizon geometry of the form AdS$_2\times S^{n}$. We then lifted 
these geometries up to ten or eleven dimensions on $S^m$, where they 
all have the form of a warped $S^n\times
 S^m$ bundle over AdS$_2$. The reparameterisations of the azimuthal 
directions of $S^m$, which are fibred over the AdS$_2$, are associated 
with Virasoro symmetries and the central charges of the corresponding 
two-dimensional chiral CFTs can be read off. All of the higher-dimensional 
geometries have the form of a manifold bundle over AdS$_2$, which we 
have shown implies that the microscopic entropy of each CFT, calculated 
using the Cardy formula, will agree precisely with the
 Bekenstein-Hawking entropy.

There are two seemingly distinct holographic dualities that can be applied to the above backgrounds, namely, the AdS/CFT correspondence and the extremal black hole/CFT correspondence. Before the near-horizon limit is taken, from the AdS/CFT perspective the asymptotically locally AdS geometry corresponds to a conformal fixed point in the UV region of a Yang-Mills theory at finite temperature. The charges of the AdS black hole are associated with the presence of chemical potentials \cite{kraus}. Coming in towards the black hole corresponds to the field theory undergoing a Renormalization Group flow. Thus, the near-horizon region corresponds to the low-energy limit of the Yang-Mills theory. At the same time, from the extremal black hole/CFT perspective, the near-horizon region corresponds to a set of two-dimensional CFTs, whose central charges are related to the charges of the AdS black hole. Via these two types of holographic dualities, the low-energy limit of the Yang-Mills theory is described by the two-dimensional CFTs. It would be interesting if this link of the triality could be made direct. However, since the near-horizon geometry contains an AdS$_2$ component, this indicates that the AdS$_2$/CFT$_1$ correspondence might apply, which is still not well understood. Thus, perhaps one type of holographic correspondence could help shed light on the other.

\section*{Acknowledgement}
The research of H.L., J.M. and C.N.P. is supported in part by DOE
grant DE-FG03-95ER40917

\end{document}